\documentclass[lettersize,journal]{IEEEtran}
\usepackage{amsmath,amsfonts}
\usepackage{algorithmic}
\usepackage{algorithm}
\usepackage{array}
\usepackage[caption=false,font=normalsize,labelfont=sf,textfont=sf]{subfig}
\usepackage{textcomp}
\usepackage{stfloats}
\usepackage{url}
\usepackage{verbatim}
\usepackage{graphicx}
\usepackage{cite}
\usepackage{color}
\usepackage{multirow}
\usepackage{rotating}
\begin{document}
\title{Subject-Independent Deep Architecture for EEG-based Motor Imagery Classification}
\author{Shadi Sartipi, \IEEEmembership{Student Member, IEEE}, and Mujdat Cetin, \IEEEmembership{Fellow, IEEE}
\thanks{This work has been partially supported by the National Science Foundation (NSF) under grants CCF-1934962 and DGE-1922591. ($corresponding~author:~Shadi~Sartipi$)}
\thanks{Shadi Sartipi and Mujdat Cetin are with the Department of Electrical and Computer Engineering, University of Rochester, Rochester, NY 14627, USA ({ssartipi@ur.rochester.edu, mujdat.cetin@rochester.edu}).}
\thanks{Mujdat Cetin is also with the Goergen Institute for Data Science, University of Rochester, Rochester, NY 14627, USA.}}

\maketitle

\begin{abstract}
Motor imagery (MI) classification based on electroencephalogram (EEG) is a widely-used technique in non-invasive brain-computer interface (BCI) systems. Since EEG recordings suffer from heterogeneity across subjects and labeled data insufficiency, designing a classifier that performs the MI independently from the subject with limited labeled samples would be desirable. To overcome these limitations, we propose a novel subject-independent semi-supervised deep architecture (SSDA). The proposed SSDA consists of two parts: an unsupervised and a supervised element. The training set contains both labeled and unlabeled data samples from multiple subjects. First, the unsupervised part, known as the columnar spatiotemporal auto-encoder (CST-AE), extracts latent features from all the training samples by maximizing the similarity between the original and reconstructed data. A dimensional scaling approach is employed to reduce the dimensionality of the representations while preserving their discriminability. Second, a supervised part learns a classifier based on the labeled training samples using the latent features acquired in the unsupervised part. Moreover, we employ center loss in the supervised part to minimize the embedding space distance of each point in a class to its center.  The model optimizes both parts of the network in an end-to-end fashion. The performance of the proposed SSDA is evaluated on test subjects who were not seen by the model during the training phase.  To assess the performance, we use two benchmark EEG-based MI task datasets. The results demonstrate that SSDA outperforms state-of-the-art methods and that a small number of labeled training samples can be sufficient for strong classification performance.
\end{abstract}


\section{Introduction}
\label{sec:introduction}
\IEEEPARstart{M}{otor} imagery (MI)  brain-computer interfaces (BCI) enable interpreting the imagination of limb movements. MI BCI has considerable biomedical applications in areas such as neuro-rehabilitation \cite{mak2009clinical}. Electroencephalogram (EEG) as a non-invasive method for recording the human brain's electrical activity has been widely used in many MI BCI contexts due to its cost-effective and non-invasive nature \cite{ang2011large}. Machine learning (ML) approaches have shown great promise in extracting meaningful information from EEG data \cite{sharma2022motor}. Nevertheless, dealing with EEG data involves challenges due to their heterogeneity across different subjects engaged in the same task \cite{suk2012novel}. Also, these recordings always carry noise and various artifacts which can negatively impact the performance of computational models. As a result, the development of effective models to analyze the imaginary limb movements continues to be an active research topic.

Several articles have been published in the literature to solve the EEG-based MI task \cite{cantillo2014approach}. In principle, two approaches can be used to tackle the aforementioned problems. In the first approach, the automated system is calibrated and trained on the specific subject and then the learned network is applied to perform the classification task on the same subject which is called subject-dependent classification. Since each subject has their individual way of reacting to mental tasks and the model is calibrated based on their own way of thinking, this technique leads to acceptable classification performance, as long as we can collect sufficient labeled data and train the classifier for each subject. Despite the effectiveness of this method, generalization to a broad population of users is not in principle guaranteed. The second approach focuses on developing a generalized system capable of application across diverse subjects engaging in a similar task, known as subject-independent classification. While this approach is more practical and desirable in many respects, applying a subject-independent model to individual subjects often results in lower classification accuracy compared to the subject-dependent approach due to individual differences among subjects. Therefore, constructing a subject-independent model that can perform well enough on new subjects would be desirable.

Recently, deep learning models have been shown to exhibit superior performance in EEG recordings compared to traditional machine learning algorithms \cite{craik2019deep}. Two main drawbacks of the majority of the existing models are the poor performance of the model in the subject-independent scenario and the model's dependency on a sufficient amount of labeled data in a purely supervised learning setting \cite{lawhern2018eegnet, zhang2018cascade}. However, labeling the EEG recordings is costly, difficult, and time-consuming \cite{zheng2018emotionmeter}.

In this paper, we propose a new approach for solving left/right-hand, foot, and tongue movement imagination tasks using a novel subject-independent semi-supervised deep architecture (SSDA). The network contains an auto-encoder (AE) which is trained in an unsupervised fashion (i.e., without labels) on training subjects. The AE learns a latent representation through the process of maximizing the similarity between original and reconstructed EEG data. The latent representation extracted by the AE is fed into a classifier trained using a small amount of labeled data in a supervised fashion. The classifier and the AE are trained simultaneously in an end-to-end manner. To address the complexity of finding the best hyper-parameters in deep architectures, we propose a columnar structure for the AE where each column consists of convolutional neural networks (CNN) and recurrent neural networks (RNN) followed by an attention model leading to a columnar spatio-temporal auto-encoder (CST-AE). The CST-AE has the ability to incorporate different spatio-temporal windows to learn the latent representations. We can find a lower dimensional representation without losing the discriminative power via dimensional scaling (DS) in the encoder part \cite{gong2019intrinsic}. Thus, the reconstruction loss consists of mean-square-error (MSE) loss and DS loss. The classifier involves both a cross-entropy loss and a  center loss \cite{ghosh2018understanding} to minimize the embedding space distance of each data sample to its class center. During the training phase, the parameters of CST-AE and the classifier are optimized by minimizing the weighted linear combination of reconstruction loss and the classifier loss. As mentioned before, we optimize the CST-AE network without labels (in an unsupervised fashion), thanks to which our proposed architecture can perform well even when the number of labeled training samples is limited. We evaluate the proposed model's performance on test subjects who have not been seen by the model during the training phase, leading to a subject-independent structure. We apply our approach to two of the publicly available datasets, namely, PhysioNet (105 subjects) \cite{goldberger2000physiobank} and BCI Competition IV 2a (9 subjects) \cite{brunner2008bci} in a subject-independent fashion, in order to validate the performance of the proposed method.

The main contributions of this work are highlighted as follows:
\begin{itemize}
    \item A novel subject-independent SSDA approach is proposed for EEG-based MI tasks. The model consists of a deep unsupervised CST-AE along with a supervised deep classifier. The CST-AE extracts the spatial, temporal, and attentive information to learn the latent representations without relying on just one fixed spatio-temporal window. 
    \item Dimensional scaling is applied in the proposed encoder part of the network to obtain the lower dimensional representations while maintaining the discriminative ability to a large extent.
    \item A new loss function is defined for the supervised classifier part of the network which not only minimizes the loss based on the given labels but also minimizes the intra-class variability. The comprehensive experiments are performed to show the significance of the defined loss function's performance with a limited number of labeled trained samples.
\end{itemize}

The rest of this paper is organized as follows. Section~\ref{sec:related} summarizes the related work. The proposed approach is described in Section~\ref{sec:method}. Details of implementation and datasets are included in Section~\ref{sec:setup}. Section~\ref{sec:experiment} presents the experimental results and discussions. Section~\ref{sec: conclusion} concludes the paper.
\section{Related Work}
\label{sec:related}
EEG-based MI classification is the basis of BCI and numerous approaches have been published \cite{hamedi2016electroencephalographic, edelman2015eeg}. Traditional algorithms commonly consist of two phases, namely, hand-crafted feature extraction and classification. The popular approach is to investigate EEG data in specific frequency bands by calculating the power spectral density (PSD) as a feature \cite{lotte2010regularizing}. Mutual-information-based features in spatial and temporal domains were also used to classify EEG data \cite{meng2014simultaneously}. Edelman \emph{et~al.} applied principal component analysis for the EEG-based MI task \cite{edelman2015eeg}. In sensorimotor rhythms, event-related synchronization and desynchronization induced by movement imagination have been widely studied in EEG signals while performing MI tasks \cite{hamedi2016electroencephalographic}. Filter bank common spatial pattern (FBCSP) is a widely used feature extraction method that applies the common spatial patterns (CSP) to different frequency bands and chooses the discriminative features in a subject-dependent fashion \cite{ang2008filter}. Gaur \emph{et~al.} \cite{gaur2021sliding} were able to perform a binary classification via two different sliding window-based CSP, where they consider multiple time segments in each trial. The sparse support matrix machine model is proposed by \cite{zheng2018sparse} to consider the structural information and feature selection at the same time in order to improve the EEG classification performance.    

Lately, deep neural networks (DNN) have been applied widely in MI classification tasks \cite{craik2019deep}. DL algorithms bring the possibility of learning the discriminative features from the raw EEG recordings in an end-to-end fashion by combining representation learning and classifier learning. In \cite{kumar2016deep}, multi-layer perceptron along CSP features is applied to replace the traditional classifiers. CNNs have been shown to be effective in encoding the spatial and structural information in EEG \cite{zhang2018cascade}. DeepConvNet \cite{schirrmeister2017deep} and EEGNet \cite{lawhern2018eegnet} are two CNN-based models which demonstrated superiority in multiple EEG-based tasks. In \cite{han2021semi}, the authors presented a domain-independent semi-supervised approach using EEGNet and DeepConvNet as backbones. CNN with multiple 1D convolutional layers is applied to raw EEG data in \cite{tang2017single}. Zhao \emph{et~al.} \cite{zhao2019multi} generated the 3D representation of the EEG data by transforming them into sequences of 2D arrays and applied multi-branch 3D CNN to extract the features. Sakhavi \emph{et~al.} altered the FBCSP to find the temporal features and applied CNN for EEG decoding \cite{sakhavi2018learning}. Ko \emph{et~al.} extracted spatio-spectral-temporal features and fed them as the input of the CNN network \cite{ko2021multi}. The network was able to learn complex representations. Another research group proposed an adaptive transfer learning strategy using CNN as a backbone of DNN \cite{zhang2021adaptive}. During the test phase, they applied a small number of test samples to calibrate the pre-trained network before performing the classification task. RNNs are known for their ability to represent temporal dynamics \cite{zhang2018cascade}. In \cite{liu2022fbmsnet} a filter bank was applied to get different spectral EEG representations and the spatial and temporal CNN was used to extract the features. To remove the noise source from EEG data, Hwaidi \emph{et~al.} \cite{hwaidi2022classification} used an auto-encoder prior to the CNN layer. 

Several subject-independent approaches have emerged to tackle the MI task. Zhang \emph{et~al.} \cite{zhang2020motor} presented the graph-based convolutional recurrent model as a deep architecture that uses the graph representation to get the spatial and temporal dynamics. The Multi-subject ensemble CNN was proposed in \cite{dolzhikova2022subject}, where K-folds were employed to train K base CNN classifiers using a held-out fold. Nagarajan \emph{et~al.} \cite{nagarajan2022relevance} explored the use of layer-wise relevance propagation and neural network pruning techniques for subject-independent channel selection, aiming to enhance decoding performance. The combination of CSP and CNN was used in \cite{nouri2023towards} to get the spatial and spectral EEG features, respectively. 

Although DL algorithms improve the classification performance, they require a large number of labeled samples to train on. Semi-supervised learning (SSL) has been widely used to overcome labeled data scarcity issues \cite{nicolas2015adaptive}. Some of the common SSL approaches are graph-based methods, expectation-maximization (EM), pseudo-labeling, generative models, $\Pi$ model, and temporal ensembling \cite{zhu2005semi, ko2022semi, zhang2021deep}. Graph-based methods propagate the limited labeled samples to the unlabeled ones which were applied in the MI task with a novel iteration mechanism \cite{guan2013joint}. EM was also used widely in BCI applications for CSP-based classification and unsupervised adjustment of Gaussian mixture models \cite{liu2010improved}. Lee \emph{et~al.} proposed the pseudo labeling approach where the network was trained on the limited labeled data and the trained network was used to produce the pseudo labels for unlabeled samples \cite{lee2013pseudo}. Then, the network was re-trained on all labeled and pseudo-labeled samples. Generative models were used to generate synthetic data and learn the distributional characteristics of EEG data \cite{ko2022semi}. The $\Pi$ model was proposed to deal with the case where the number of labeled samples was limited enough making the pseudo labeling unstable \cite{oliver2018realistic}. The idea was to add noise and dropout on the input data and network, respectively. Then, two different outputs would be obtained for each individual input sample and the network would seek to minimize the distance between the two outputs. This would force the network to make equivalent predictions for the same input with different additive noises \cite{oliver2018realistic}. To improve the training speed of the $\Pi$ model, temporal ensembling was introduced which aggregates the network output from previous epochs into an ensemble \cite{laine2016temporal}. 
\begin{figure}[t!]
    \centering
    \includegraphics[width=1\linewidth]{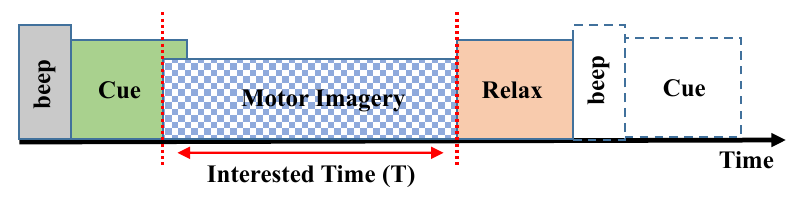}
    \caption{Motor imagery EEG acquisition experiment.}
    \label{fig0}
\end{figure}
\begin{figure}[t!]
    \centering
    \includegraphics[width=1\linewidth]{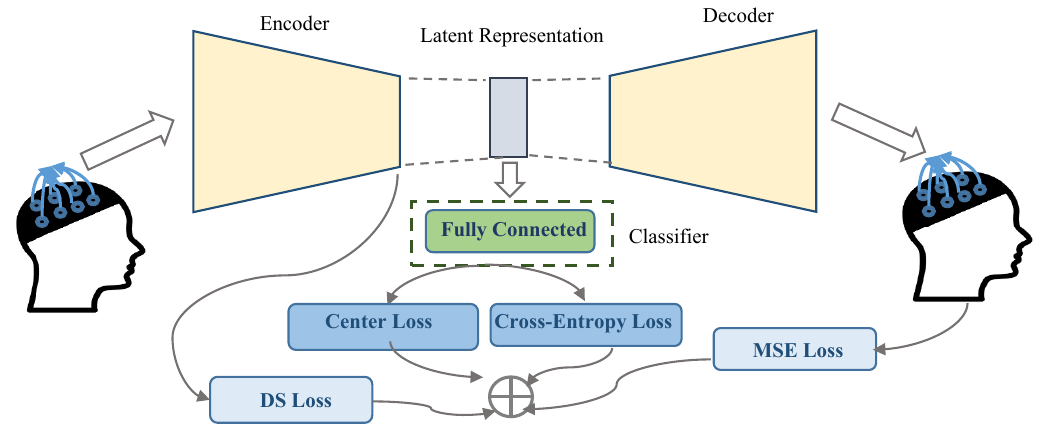}
    \caption{Block-diagram of the proposed subject-independent semi-supervised deep architecture.}
    \label{fig1}
\end{figure}
\begin{figure*}[t!]
    \centering
    \includegraphics[width=1\linewidth]{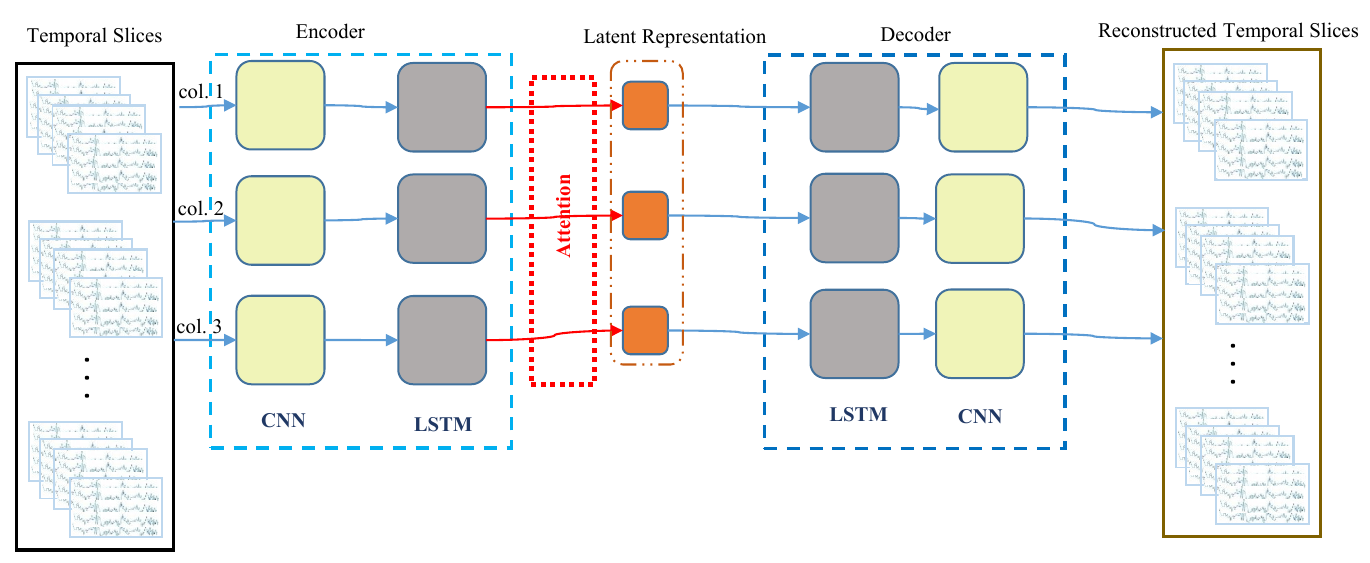}
    \caption{Proposed columnar spatio-temporal auto-encoder (CST-AE) architecture. Yellow and grey blocks represent CNN and LSTM layers, respectively. Orange blocks are the latent representations that are the outputs of the attention mechanism. (col: Column).}
    \label{ae}
\end{figure*}
\begin{table*}[t]
\centering
\caption{Parameters of the layers used in the Columnar Spatio-Temporal Auto-Encoder. $C$ and $U$ are the number of electrodes and row-wise upsampling, respectively. (col.: column, Pool: maxpooling, CNN: 2D CNN, Sample: upsampling, BN: Batch normalization).}

\begin{tabular}[t]{ c c c c c c|c c c c c c}
\hline
 \multicolumn{6}{c}{Encoder$^{*}$} & \multicolumn{6}{c}{Decoder$^{**}$} \\

\hline
\hline
\multicolumn{1}{c}{col. 1} & \multicolumn{1}{c|}{}&\multicolumn{1}{c}{col. 2}& \multicolumn{1}{c|}{} &\multicolumn{1}{c}{col.3}& \multicolumn{1}{c|}{}& \multicolumn{1}{c}{col.1} & \multicolumn{1}{c|}{}&  \multicolumn{1}{c}{col.2} & \multicolumn{1}{c|}{}& \multicolumn{1}{c}{col.3}& \multicolumn{1}{c}{}\\
\hline
\multicolumn{1}{c}{layer}&\multicolumn{1}{c|}{parameter} & \multicolumn{1}{c}{layer} &  \multicolumn{1}{c|}{parameter} & \multicolumn{1}{c}{layer}&\multicolumn{1}{c|}{parameter}&\multicolumn{1}{c}{layer}&\multicolumn{1}{c|}{parameter} & \multicolumn{1}{c}{layer} &  \multicolumn{1}{c|}{parameter} & \multicolumn{1}{c}{layer}&\multicolumn{1}{c}{parameter}\\
\hline
CNN&$(64;C,50)$&CNN&$(40;C,45)$&CNN&$(30;C,15)$&LSTM&$(100;0.2)$&LSTM&$(40;0.4)$&LSTM&$(30;0.2)$\\
BN&$-$&BN&$-$&BN&$-$&Reshape&$(1,2,50)$&Reshape&$(1,2,20)$&Reshape&$(1,2,15)$\\
Pool&$(1,80)$&Pool&$(1,75)$&Pool&$(1,35)$&Sample&$(U,4)$&Sample&$(U,4)$&Sample&$(U,4)$\\
Dropout&$0.5$&Dropout&$0.5$&Dropout&$0.5$&BN&$-$&BN&$-$&BN&$-$\\
Flatten&$-$&Flatten&$-$&Flatten&$-$&CNN&$(64;7,7)$&CNN&$(40;7,7)$&CNN&$(30;7,7)$\\
LSTM&$(64;0.4)$&LSTM&$(40;0.4)$&LSTM&$(30;0.2)$&BN&$-$&BN&$-$&BN&$-$\\
{}&{}&{}&{}&{}&{}&CNN&$(1;1,1)$&CNN&$(1;1,1)$&CNN&$(1;1,1)$\\
\hline

\hline
\end{tabular}
\begin{tabular}[t]{lc}
*parameters format: CNN (number of filters; filter size) with Relu activation and valid padding, Pool (pool size), Dropout (dropout rate), LSTM (filter size,\\
dropout rate). \\
**parameters format: CNN (number of filters; filter size) with Relu activation and same padding, Sample (row-wise scale, horizontal scale), LSTM (filter size,\\ dropout rate).\\
\end{tabular}
\label{table1}
\end{table*}

\section{Proposed Approach}
\label{sec:method}
To begin, let us consider a typical timing scheme for an MI experiment, as depicted in Fig.~\ref{fig0}. The beep and cue are utilized to alert the subject of the trial's start time and prompt them to engage in the MI task. As illustrated in Fig.~\ref{fig0}, during each trial, the focus is on the interested time denoted as T. T represents the MI engagement task, starting after a couple of milliseconds of the appearance of the cue and lasting until the relaxation period, as indicated by the disappearance of the fixation cross. Subjects perform the MI task during this T period, followed by a brief relaxation period as the screen returns to normal.

In this section, we explain the proposed subject-independent semi-supervised deep architecture for EEG representation learning and classification (Fig.~\ref{fig1}). First, we explain the proposed columnar spatio-temporal auto-encoder (CST-AE). Second, we describe the classifier. Finally, the semi-supervised procedure is explained.
\subsection{Columnar Spatio-Temporal Auto-Encoder}
EEG data are recorded over the electrodes which can be shown as $Z\in\mathbb{R}^{C\times{}T}$, where $C$ is the number of the EEG electrodes, and $T$ represents the number of time points (interest time). Sliding window with length $m$ with an overlap $p$ is applied on raw EEG data to get the temporal time slices $D_i\in\mathbb{R}^{C\times{}m}$, where $m$ is the temporal slice length, $i={1, 2,...,n}$, and $n=\text{floor}((T-m)/p)+1$. For the rest of the paper, we consider $l$ and $u$ as the sets of labeled and unlabeled data samples, respectively. We use $N$ to denote the total number of training data samples, $N_l$ to denote the total number of labeled data samples, and $N_u$ to denote the total number of unlabeled data samples. We assume that $l \cap u = \emptyset$.  

The proposed CST-AE is a columnar auto-encoder \cite{li2020extraction} including the encoder that maps the input EEG data with $N$ samples into a latent space, and a decoder to reconstruct the input from the latent variables. The architecture is illustrated in Fig.~\ref{ae}. The encoder consists of a CNN layer to learn structural and spatial representations and an LSTM layer followed by an attention mechanism to learn temporal representations and attentive information. Since finding the best kernel sizes as model parameters associated with the CNN and LSTM layers is challenging, we design the model in a columnar fashion. In each column the EEG slices are encoded by a $2$ dimensional (2D) spatial CNN to get the higher-order representation$\{S_i|S_i=\text{Conv}(D_i),~i=[1,n]\}$. A rectified linear unit (ReLU) activation function and a valid padding option are used in the convolution encoding layers. The output is fed to a maxpooling layer to obtain
\begin{equation}
    Q_i=\text{MaxPool}(S_i),~i=[1,n]
\end{equation}  
The output of the last layer is flattened and that leads to $n$ $1$D feature vectors. One layer of LSTM with an attention mechanism is applied to get the informative temporal dynamics. Thus, the output of each LSTM cell would be $\{h_{i}^{enc}|h_{i}^{enc}=\text{LSTM}(Q_i),~i=[1,n]\}$.
The attention mechanism is employed to emphasize the temporal slices that the subject pays attention to during task performance. Considering $W$ and $b$ as trainable weights and biases, respectively, the attention weights and the output of the attention mechanism, $\alpha_i$ and $v$,  are calculated as follows \cite{sartipi2023hybrid}: 
\begin{equation}
    v=\sum_{i}{\alpha_{i} h_{i}^{enc}}, \;\; \alpha_{i}= \frac{\exp(W h_{i}^{enc}+b)}{\sum_{j}\exp(W h_{i}^{enc}+b)}.
\end{equation}

The decoder part contains an LSTM and two CNN layers. First, the latent variable, $v$, is fed to the LSTM cell as shown in Eq. $3$. Second, the result is upsampled and fed to the first CNN layer (Eq. $4$). Finally, a CNN layer with kernel size $1$ is applied to get the reconstruction, $\hat{D}_i$, as shown in Eq. $5$.
\begin{equation}
    h_{i}^{dec}=\text{LSTM}(v),~i=[1,n]
\end{equation}
\begin{equation}
\tilde{D}_i=\text{Conv}(\text{UpSample}(h_{i}^{dec})),~i=[1,n]
\end{equation}
\begin{equation}
    \hat{D}_i=\text{Conv}(\tilde{D}_i),~i=[1,n]
\end{equation}
Table~\ref{table1} shows the set of parameters and implementation details for CST-AE.

In addition, we utilize dimensional scaling (DS) \cite{kruskal1964multidimensional} in the last layer of the encoder to reduce the dimensionality of the extracted features while preserving their discriminative ability. To this end, we define $Z$ as the original EEG data and $V$ as the corresponding extracted features in the attention layer. The DS is formulated as an optimization problem that minimizes the squared difference between the similarity indicators of $Z$ and $V$, i.e.,
\begin{equation}
\mathbf{minimize}~\sum_{i=1}^{N}\sum_{j=1}^{N}[d_{H}(Z_i,~Z_j)-d_{L}(V_{i},~V_{j})]^2,
\end{equation}
where $d_H$ and $d_L$ represent similarity indicators \cite{gong2019intrinsic} between the original EEG data and lower-dimensional representations in the attention layer, respectively.
\subsection{Classifier}
As shown in Fig.~\ref{fig1}, the classifier block is defined to perform supervised classification on labeled training samples, $l$. The latent representation obtained from the encoder is fed to the classifier, which contains two fully connected (FC) layers with $128$ and $2$ units, respectively. $L2$ kernel regularization \cite{cortes2012l2} with factor $0.0005$ is applied to the second FC layer.  
\subsection{Semi-Supervised Deep Architecture}
Two different loss functions are defined, namely, the unsupervised reconstruction loss, $\mathcal{L}_{un}$, and the supervised classification loss, $\mathcal{L}_s$. In the supervised part of the architecture, $\mathcal{L}s$ consists of two parts, namely cross-entropy loss, $\mathcal{L}{ce}$, and center loss, $\mathcal{L}_c$, defined as below:

\begin{equation}
\mathcal{L}{ce}=-\frac{1}{N_l}\sum{i=1}^{N_l}y_i\text{ln}(y_i^{\prime})
\end{equation}

\begin{equation}
\mathcal{L}{c}=-\frac{1}{2}\sum{i=1}^{N_l}|f(D_i,\theta)-c_{y_i}|_2^2
\end{equation}

\begin{equation}
\mathcal{L}s=\mathcal{L}{ce}+\gamma\mathcal{L}_{c}
\end{equation}
Here, $y$ and $y^{\prime}$ are the actual and predicted labels, respectively. $f(-,\theta)$ denotes the parametric function for latent variable calculation with parameter $\theta$. $c_{y}$ is the $y{\text{th}}$ target class center, and $\gamma$ is the constant weight. The center loss would be helpful in minimizing intra-class variations.

$\mathcal{L}_{un}$ consists of a mean-square-error (MSE) loss, $\mathcal{L}_{mse}$, to minimize the differences between the input and the reconstructed input, and a DS loss, $\mathcal{L}_{ds}$, to keep the model discriminative to a large extent. $\mathcal{L}_{mse}$ is defined as follows: 
\begin{equation}
    \mathcal{L}_{mse}=\sum_{i=1}^{M}\beta_{i}(\sum_{j=1}^{N_l}\|D_j -\hat{D}_j\|_2^2+\sum_{j=1}^{N_u}\|D_j^{\prime} -\hat{D}_j^{\prime}\|_2^2)
\end{equation}
\begin{equation}
\begin{aligned}[b]
    \mathcal{L}_{ds}=\sum_{i=1}^{M}\eta_{i}&(\sum_{k=1}^{N_l}\sum_{l=1}^{N_l}[d_{H}(D_k,~D_l)-d_{L}(V_{k},~V_{l})]^2+\\
    &\sum_{k=1}^{N_u}\sum_{l=1}^{N_u}[d_{H}(D_k^{\prime},~D_l^{\prime})-d_{L}(V_{k}^{\prime},~V_{l}^{\prime})]^2)
    \end{aligned}
\end{equation}
\begin{equation}
    \mathcal{L}_{un}=\mathcal{L}_{mse}+\mathcal{L}_{ds}
\end{equation}
where $M$ is the number of the columns in CST-AE, $D,~D^{\prime}$ denote the labeled and unlabeled input data samples, respectively, $\hat{D},~\hat{D}^{\prime}$ correspond to the reconstructed labeled and unlabeled data samples, respectively, and $V,~V^{\prime}$ denote the latent variables of labeled and unlabeled input data, $V=f(D,\theta)$ and $V^{\prime}=f(D^{\prime},\theta)$, respectively. Also, $d_H,~d_L$ are Euclidean distances in original data and lower dimension spaces. $\beta$ and $\eta$ are constant weights that enable curriculum learning.

Taking both supervised and supervised components into account jointly, the final loss function is calculated as below:
\begin{equation}
    \mathcal{L}=\mathcal{L}_{un}+\mathcal{L}_s
\end{equation}
Grid search in range $[0,~0.5]$ with step-size $0.1$ is performed to get the best values for $\beta$, $\eta$, and $\gamma$.
 \begin{table}[t!]
\centering
\caption{Classification performance of the proposed SSDA on each validation fold (V) when $N_l=N$.}
{\begin{tabular}{clllclc}
\hline
\textbf{Dataset} &\textbf{}& \textbf{Fold} &\textbf{}& \textbf{Accuracy}&\textbf{} &  \textbf{F1 Score} \\
\hline\hline
\multirow{10}{*}{\begin{sideways}PhysioNet\end{sideways}}& & V1 & & $0.81$& & $0.80$\\ 
 & &  V2&  & $0.84$& & $0.83$ \\ 
 &  &V3&  & $0.86$& & $0.86$ \\ 
 &  &V4&  & $0.87$& & $0.87$\\ 
 & &V5& & $0.81$& & $0.80$\\
 & &V6&  & $0.85$& & $0.84$\\
 & &V7&  & $0.84$& & $0.84$\\
 & &V8&  & $0.77$& & $0.77$\\
 & &V9&  & $0.84$& & $0.84$\\
 & &V10&  & $0.81$& & $0.81$\\\hline \hline \hline
 \multirow {9}{*}{\begin{sideways} BCI IV 2a\end{sideways}} & &V1&  & $0.58$& & $0.58$\\ 
 &  &V2&  & $0.68$& & $0.68$ \\ 
 &  &V3&  & $0.68$& & $0.67$ \\ 
 &  &V4&  & $0.47$& & $0.45$\\ 
 & &V5& & $0.71$& & $0.70$\\
 & &V6& & $0.62$& & $0.59$\\
 & &V7& & $0.51$& & $0.50$\\
 & &V8& & $0.68$& & $0.66$\\
 & &V9& & $0.53$& & $0.53$\\\hline

\end{tabular}}
\label{classification}
\end{table}
\section{Experimental Setup}
\label{sec:setup}
\subsection{Dataset}
In order to evaluate the proposed method's performance, we used two publicly available MI benchmarks: PhysioNet MI EEG dataset \cite{goldberger2000physiobank} and BCI Competition IV 2a \cite{brunner2008bci}. These benchmarks include two-class and four-class MI classification tasks, respectively.
\subsubsection{PhysioNet Dataset} This dataset includes EEG recordings of $109$ healthy subjects. During the experiment, a target appears either on the left or right side of the screen. The participant imagines opening and closing the corresponding fist until the target vanishes, followed by a period of relaxation. BCI$2000$ instrument with $64$ EEG electrodes is used for data collection. The sampling rate is set to $160$ Hz and each trial lasts for $3.1$ seconds. The recordings related to subjects $88$,~$89$,~$92$, and $100$ are removed due to technical problems and large amounts of rest periods \cite{goldberger2000physiobank}. This resulted in $105$ subjects and each subject performed $45$ trials roughly.  For evaluation, we used $10$-fold cross-validation with 10 repetitions. In each repetition, ten randomly selected subjects were used as the test set, and the remaining subjects were used as the training set. The final performance was calculated as the average performance across all repetitions.
\subsubsection{BCI Competition IV 2a Dataset} This dataset comprises EEG recordings from $9$ healthy participants. The cues related to the BCI paradigm correspond to four classes, namely the imagination of movement of the tongue, feet, right hand, and left hand. EEG data are recorded over $22$ electrodes with a $250$ Hz sampling frequency. Each subject performs the four-class task in two sessions on two different days; a training session and a testing session, respectively. Each session consists of $288$ EEG trials, each of which $4$ s. Cross-validation was performed using the leave-one-subject-out method. This means that both the training and testing sessions of one subject were used as a test set, while all sessions of the other subjects were used as a training set each time. 
\subsection{Implementation Details}
Each EEG trial is sliced into temporal fragments. With datasets having sampling frequencies of $160$ and $250$ Hz, we have chosen the same window length, $m$, of $400$ samples with different step sizes, $p$. This duration allows sufficient time for the brain to initiate motor imagery execution. $p$ is set to $20$ and $50$ for the PhysioNet and BCI \uppercase\expandafter{\romannumeral4} 2a, respectively. Since each dataset has a different number of electrodes, the values for $(C,U)$ pairs mentioned in Table~\ref{table1}, are set to $(64,4)$ and $(22,2)$ for PhysioNet and BCI \uppercase\expandafter{\romannumeral4} 2a, respectively. Implementation is done in Python with Tensorflow $2.8.2$. 

To optimize the model parameters and avoid over-fitting during the training process, $10\%$ of the training data samples are randomly selected as the validation data. The model parameters with the highest validation classification accuracy are considered the final trained model.  The final model is applied to the test set to get the classification performance. The number of epochs and the initial learning rate are set to $250$ and $0.00001$, respectively. Adam optimizer with default learning-rate decay is used for the optimization process. The $\mathbf{\beta}=[\beta_1,\beta_2,\beta_3]$, $\mathbf{\eta}=[\eta_1,\eta_2,\eta_3]$, and $\gamma$ values corresponded to the highest classification performance on validation data when $N_l=N$, and were set to $[0.2,0.1,0.2]$, $[0.1,0.1,0.1]$, and $0.3$, respectively.
\begin{table}[t]
\centering
\caption{Comparison of the proposed SSDA with state-of-the-art methods from recent literature when all labels of the training samples are available ($N_l=N$).}
\resizebox{0.95\linewidth}{!}{\begin{tabular}{cccc}
\hline
\textbf{Dataset} & \textbf{Method} & \textbf{Accuracy} &  \textbf{F1 Score} \\
\hline\hline
\multirow{7}{*}{\begin{sideways}PhysioNet\end{sideways}} & FBCSP \cite{ang2008filter} & $0.59\pm0.03$ & $0.60\pm0.03$\\ 
 &  RCNN \cite{bashivan2015learning} & $0.57\pm0.02$ & $0.57\pm0.01$ \\ 
 &  EEGNet \cite{lawhern2018eegnet} & $0.72\pm0.04$ & $0.72\pm0.03$ \\ 
 &  CasCNN \cite{zhang2019making} & $0.63\pm0.04$ & $0.63\pm0.03$\\ 
 & DG-HAM \cite{zhang2019graph} & $0.76\pm0.02$ & $0.77\pm0.02$\\
 & EEG-ARNN \cite{sun2022graph} & $0.82\pm0.04$ & $0.82\pm0.04$\\\cline{2-4}
 & {\bf Proposed SSDA}               & ${\bf0.83\pm0.03 }$ & ${\bf0.83\pm0.03}$\\\hline \hline \hline
 \multirow {7}{*}{\begin{sideways} BCI IV 2a\end{sideways}} & FBCSP \cite{ang2008filter} & $0.36\pm0.08$ & $0.36\pm0.10$\\ 
 &  RCNN \cite{bashivan2015learning} & $0.33\pm0.04$ & $0.33\pm0.01$ \\ 
 &  EEGNet \cite{lawhern2018eegnet} & $0.51\pm0.05$ & $0.49\pm0.03$ \\ 
 &  CasCNN \cite{zhang2019making} & $0.32\pm0.04$ & $0.32\pm0.03$\\ 
 & DG-HAM \cite{zhang2019graph} & $0.59\pm0.10$ & $0.58\pm0.10$\\
 & GSAN \cite{li2023subdomain} & $0.43\pm0.09$ & $-$
 \\\cline{2-4}
 & {\bf Proposed SSDA}               & ${\bf0.61\pm0.08 }$ & ${\bf0.59\pm0.08}$\\\hline

\end{tabular}}
\label{comparison}
\end{table}

\begin{table}[t]
\centering
\caption{Results on test set when the number of labeled training data samples are limited ( $N_l\ll N$).}
\begin{tabular}{llccc}
\hline
\textbf{Dataset} & \textbf{Performance}& $N_l=3\%N$&$N_l=10\%N$ &$N_l=30\%N$\\
\hline\hline
\multirow{4}{*}{\begin{sideways}PhysioNet\end{sideways}} &  &$N_l=124$&$N_l=409$ &$N_l=1226$\\
&  & $N_u=3961$&$N_u=3676$ &$N_u=2859$\\
& Accuracy &$0.53\pm0.04$ &$0.78\pm0.03$&$0.80\pm0.03$\\ 
 &  F1 Score & $0.42\pm0.03$& $0.77\pm0.03$& $0.80\pm0.01$\\ 
\hline \hline \hline
\multirow{4}{*}{\begin{sideways}BCI IV 2a\end{sideways}} &  &$N_l=139$&$N_l=461$ &$N_l=1383$\\
&  & $N_u=4469$&$N_u=4147$ &$N_u=3225$\\
& Accuracy &$0.30\pm0.03$ &$0.48\pm0.05$&$0.55\pm0.05$\\ 
 &  F1 Score &$0.22\pm0.02$& $0.45\pm0.04$&$0.53\pm0.03$ \\  
\hline
\end{tabular}
\label{sample}
\end{table}
\begin{figure}[t!]
    \centering
    \begin{minipage}{\linewidth}
    \includegraphics[width=0.45\linewidth]{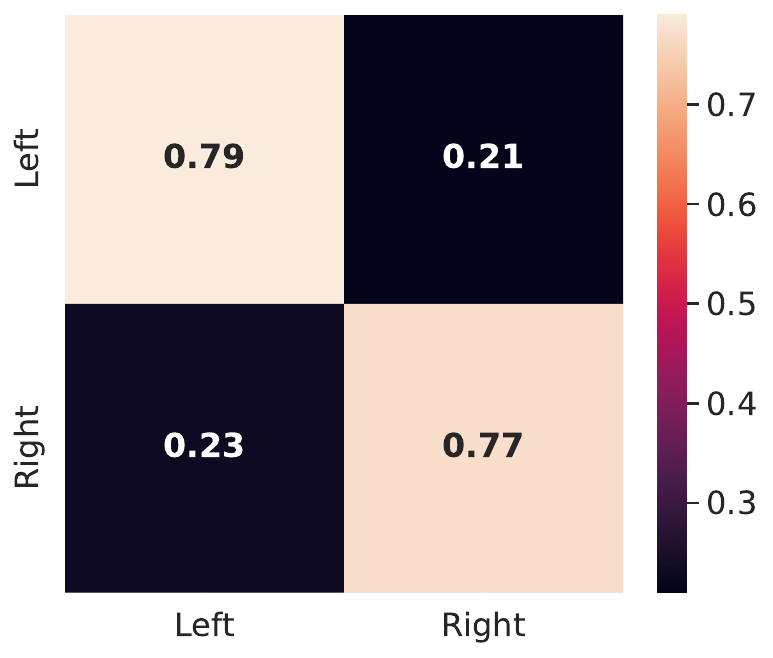}
    \includegraphics[width=0.45\linewidth]{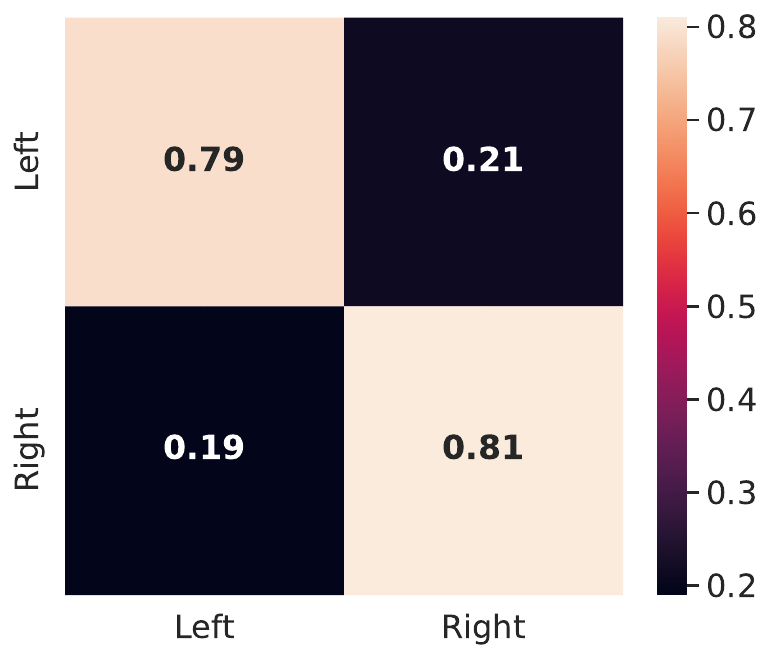}
    \end{minipage}    
    \caption{Normalized confusion matrix when $N_l\ll N$, (\textbf{left}) $10\%$ and (\textbf{right}) $30\%$ labeled training data samples on the two-class PhysioNet dataset.}
    \label{confusion1}
\end{figure}
\begin{figure}[t!]
    \centering
    \begin{minipage}{\linewidth}
    \includegraphics[width=0.45\linewidth]{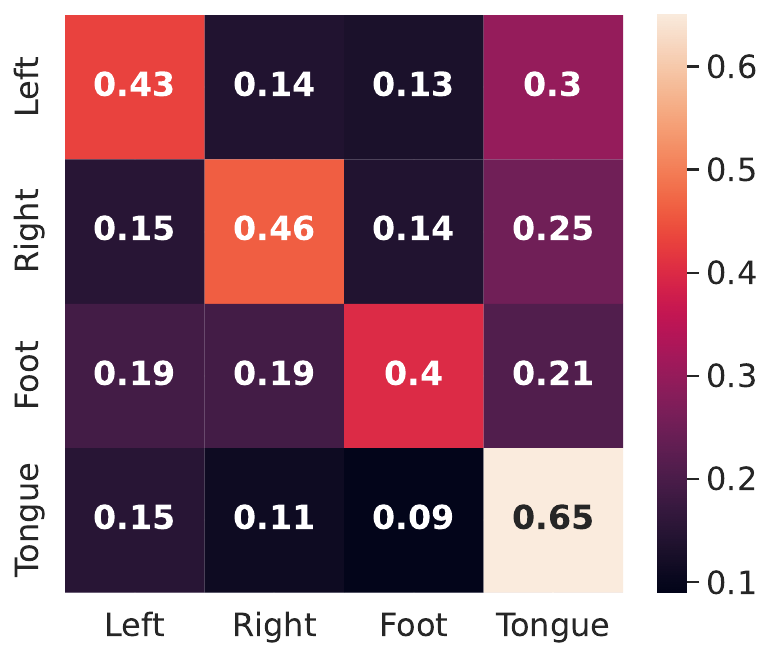}
    \includegraphics[width=0.45\linewidth]{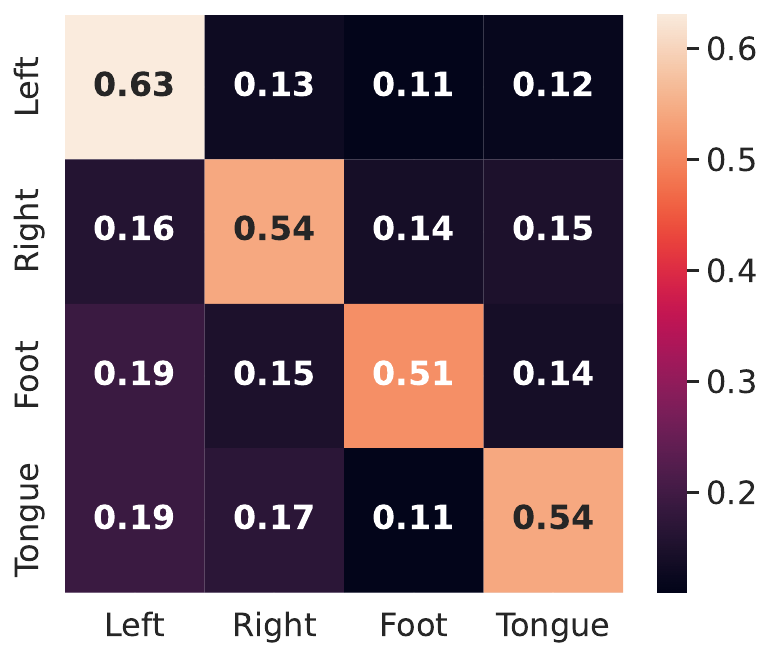}

    \end{minipage}
    \caption{Normalized confusion matrix when $N_l\ll N$, (\textbf{left}) $10\%$ and (\textbf{right}) $30\%$ labeled training data samples on the four-class BCI competition IV 2a dataset.}
    \label{confusion2}
\end{figure}
\section{Results and Discussions}
\label{sec:experiment}
\subsection{Experimental Results}
Table~\ref{classification} presents the classification accuracy for each fold when all labels of the training samples are available, $N_l=N$. As shown in Table~\ref{classification}, using raw EEG data as the input of subject-independent SSDA results in $0.83$ and $0.61$ classification accuracies for PhysioNet and BCI \uppercase\expandafter{\romannumeral4} 2a datasets, respectively.

\subsubsection{Comparison results when $N_l=N$} Since the datasets used in this paper are roughly balanced, we consider the F1 score along with classification performance. Table~\ref{comparison} summarizes the comparison results of the proposed method with state-of-the-art approaches. To have a fair comparison, all the mentioned works follow the same EEG data partitioning. First, we compare our proposed method with a traditional FBCSP \cite{ang2008filter} approach which is a common feature extraction approach that applies CSP in different frequency bands. SVM is used for classification. We also compare our work with several well-known deep learning approaches. The first deep learning approach is RCNN \cite{bashivan2015learning} which utilizes spatial, temporal, and spectral information via temporal CNN and LSTM networks. Then, we compare our work with the widely-used EEGNet \cite{lawhern2018eegnet} which is a CNN-based network. CNN blocks of the network consist of depth-wise and separable convolution operations. Furthermore, CasCNN \cite{zhang2019making}, a CNN and RNN-based model that preserves the spatio-temporal representation of the EEG data, is considered for comparison. DG-HAM \cite{zhang2019graph} is also used for comparison. This method uses graph representations and attention mechanism to perform the classification task. EEG-ARNN \cite{sun2022graph} is a graph convolutional based network that seeks to find the correlation of signals in the temporal and spatial domains. GSAN \cite{li2023subdomain} is an adversarial network that aims to detect domain-invariant features to improve subject-independent classification performance. Because the original papers for EEG-ARNN and GSAN either lack results for both datasets or employ a different cross-validation process from ours, we cannot report comparison results for both datasets. The experimental results indicate that the proposed SSDA achieves the best average classification accuracy and F1 score on both datasets.
\subsubsection{The performance of SSDA when $N_l\ll N$} As mentioned earlier, one of the major motivations of the proposed framework is dealing with a common EEG challenge namely, a limited number of labeled training samples. To explore the robustness of the proposed approach, we perform experiments where a small number of training samples are labeled ($N_l\ll N$). We randomly select $3\%$, $10\%$, and $30\%$ of the training samples as labeled samples and perform the same aforementioned classification procedure in each partition. The results for each scenario are presented in Table~\ref{sample}. Considering two class MI classification and PhysioNet dataset, the classification accuracies are $0.53\pm0.04$, $0.78\pm0.03$, and $0.80\pm0.03$ when $3\%$, $10\%$, and $30\%$ of training samples are labeled. Four class task and BCI \uppercase\expandafter{\romannumeral4} 2a reaches to $0.30\pm0.03$, $0.48\pm0.05$, and $0.55\pm0.05$ when $3\%$, $10\%$, and $30\%$ of training samples are labeled.

The normalized confusion matrices produced by SSDA with $10\%$ and $30\%$ labeled training samples for two class and four class problems are presented in Fig.~\ref{confusion1} and Fig.~\ref{confusion2}, respectively. Row labels and column labels refer to the ground truth and predicted labels, respectively.
\begin{table}[t!]
\centering
\caption{Ablation Study on the deep backbone.($N_l=N$)}
\resizebox{0.95\linewidth}{!}{\begin{tabular}{cccc}
\hline
\textbf{Dataset} & \textbf{Method} & \textbf{Accuracy} &  \textbf{F1 Score} \\
\hline
\multirow{7}{*}{\begin{sideways}PhysioNet\end{sideways}} & CNN & $0.75\pm0.03$ & $0.74\pm0.03$\\ 
 &  LSTM & $0.55\pm0.01$ & $0.53\pm0.02$ \\ 
 &  Attention  & $0.73\pm0.02$ & $0.73\pm0.02$ \\ 
 &  CNN-Attention & $0.79\pm0.03$ & $0.78\pm0.03$\\ 
 & LSTM-Attention& $0.75\pm0.09$ & $0.75\pm0.09$\\
 & CNN-LSTM &$0.73\pm0.05$ & $0.73\pm0.02$\\
 \cline{2-4}
 & {\bf Proposed SSDA}               & ${\bf0.83\pm0.03 }$ & ${\bf0.83\pm0.03}$\\\hline \hline \hline
 \multirow {7}{*}{\begin{sideways} BCI \uppercase\expandafter{\romannumeral4} 2a\end{sideways}} & CNN & $0.31\pm0.04$ & $0.29\pm0.04$\\ 
 &  LSTM & $0.33\pm0.01$ & $0.33\pm0.01$ \\ 
 &  Attention  & $0.28\pm0.02$ & $0.28\pm0.02$ \\ 
 &  CNN-Attention & $0.53\pm0.09$ & $0.52\pm0.10$\\ 
 & LSTM-Attention& $0.35\pm0.02$ & $0.35\pm0.02$\\
 & CNN-LSTM &$0.56\pm0.11$ & $0.55\pm0.12$\\
 \cline{2-4}
 & {\bf Proposed SSDA}               & ${\bf0.61\pm0.08 }$ & ${\bf0.59\pm0.08}$\\\hline

\end{tabular}}
\label{ablation}
\end{table}
\begin{table}[t!]
\centering
\caption{Ablation Study about the effectiveness of the semi-supervised learning with various deep network components.($N_l=10\% N$)}
\resizebox{0.95\linewidth}{!}{\begin{tabular}{cccc}
\hline
\textbf{Dataset} & \textbf{Method} & \textbf{Accuracy} &  \textbf{F1 Score} \\
\hline
\multirow{7}{*}{\begin{sideways}PhysioNet\end{sideways}} & CNN & $0.69\pm0.02$ & $0.69\pm0.02$\\ 
 &  LSTM & $0.55\pm0.04$ & $0.53\pm0.04$ \\ 
 &  Attention  & $0.55\pm0.05$ & $0.53\pm0.06$ \\ 
 &  CNN-Attention & $0.73\pm0.03$ & $0.73\pm0.03$\\ 
 & LSTM-Attention& $0.60\pm0.05$ & $0.59\pm0.06$\\
 & CNN-LSTM &$0.72\pm0.05$ & $0.72\pm0.04$\\
 \cline{2-4}
 & {\bf Proposed SSDA}               & ${\bf0.78\pm0.03 }$ & ${\bf0.77\pm0.03}$\\\hline \hline \hline
 \multirow {7}{*}{\begin{sideways} BCI IV 2a\end{sideways}} & CNN & $0.28\pm0.03$ & $0.26\pm0.04$\\ 
 &  LSTM & $0.25\pm0.03$ & $0.24\pm0.02$ \\ 
 &  Attention  & $0.29\pm0.03$ & $0.28\pm0.03$ \\ 
 &  CNN-Attention & $0.28\pm0.03$ & $0.27\pm0.03$\\ 
 & LSTM-Attention& $0.29\pm0.01$ & $0.28\pm0.042$\\
 & CNN-LSTM &$0.29\pm0.03$ & $0.26\pm0.03$\\
 \cline{2-4}
 & {\bf Proposed SSDA}               & ${\bf0.48\pm0.05 }$ & ${\bf0.45\pm0.04}$\\\hline

\end{tabular}}
\label{semisupervised}
\end{table}
\begin{table}[t!]
\centering
\caption{Statistical significance of the performance improvements provided by the proposed method over other methods considered in the ablation study. Wilcoxon signed-rank test p-value.($N_l=10\% N$)}
\resizebox{0.75\linewidth}{!}{\begin{tabular}{ccc}
\hline
\textbf{Method} & \textbf{PhysioNet} & \textbf{BCI IV 2a} \\
\hline
CNN & $0.002$ & $0.004$\\ 
 LSTM & $0.002$ & $0.004$ \\ 
 Attention  & $0.002$ & $0.004$ \\ 
 CNN-Attention & $0.002$ & $0.004$\\ 
 LSTM-Attention& $0.002$ & $0.004$\\
 CNN-LSTM &$0.002$ & $0.004$\\
 \hline
\end{tabular}}
\label{pvalue}
\end{table}

\begin{figure*}[t!]
    \centering
    \begin{minipage}[c]{0.24\linewidth}
    \includegraphics[width=\linewidth]{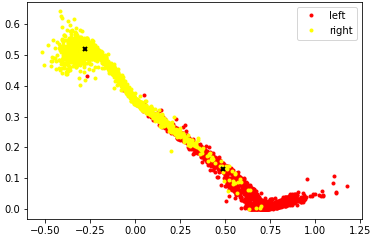}\\
    \centering{(a)}
    \end{minipage}
    \begin{minipage}[c]{0.24\linewidth}
    \includegraphics[width=\linewidth]{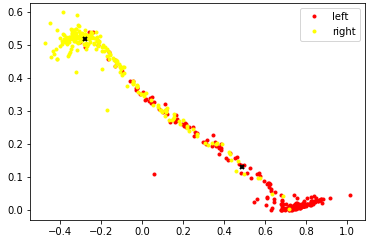}\\
    \centering{(b)}
    \end{minipage}
    \begin{minipage}[c]{0.24\linewidth}
    \includegraphics[width=\linewidth]{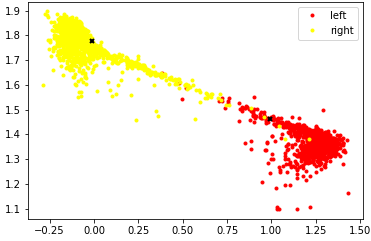}\\
    \centering{(c)}
    \end{minipage}
    \begin{minipage}[c]{0.24\linewidth}
    \includegraphics[width=\linewidth]{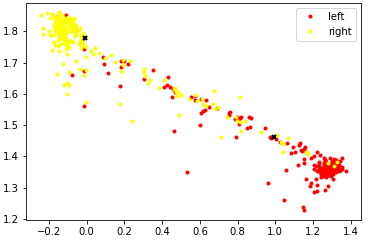}\\
    \centering{(d)}
    \end{minipage}   
    \caption{The distribution of the learned features in the last fully connected layer during (a) training phase with $\mathcal{L}_{s}=\mathcal{L}_{ce}$, (b) testing phase with $\mathcal{L}_{s}=\mathcal{L}_{ce}$, (c) training phase with $\mathcal{L}_{s}=\mathcal{L}_{ce}+\mathcal{L}_{c}$, and (d) testing phase with $\mathcal{L}_{s}=\mathcal{L}_{ce}+\mathcal{L}_{c}$ for the PhysioNet dataset. ($N_l=N$).}
    
    \label{feature}
\end{figure*}
\begin{figure*}[t!]
    \centering
    \begin{minipage}[c]{0.24\linewidth}
    \includegraphics[width=\linewidth]{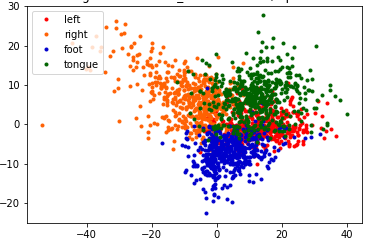}\\
    \centering{(a)}
    \end{minipage}
    \begin{minipage}[c]{0.24\linewidth}
    \includegraphics[width=\linewidth]{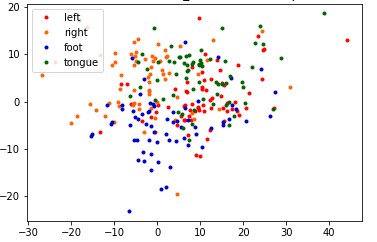}\\
    \centering{(b)}
    \end{minipage}
    \begin{minipage}[c]{0.24\linewidth}
    \includegraphics[width=\linewidth]{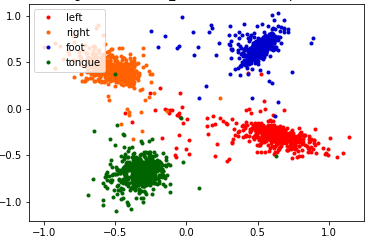}\\
    \centering{(c)}
    \end{minipage}
    \begin{minipage}[c]{0.24\linewidth}
    \includegraphics[width=\linewidth]{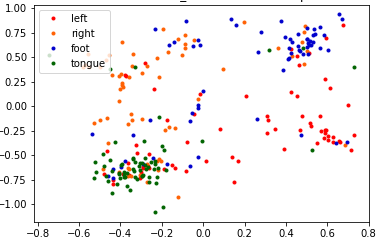}\\
    \centering{(d)}
    \end{minipage}
    
    \caption{The distribution of the learned features in the last fully connected layer during (a) training phase with $\mathcal{L}_{s}=\mathcal{L}_{ce}$, (b) testing phase with $\mathcal{L}_{s}=\mathcal{L}_{ce}$, (c) training phase with $\mathcal{L}_{s}=\mathcal{L}_{ce}+\mathcal{L}_{c}$, and (d) testing phase with $\mathcal{L}_{s}=\mathcal{L}_{ce}+\mathcal{L}_{c}$ for the BCI Competition IV 2a dataset. ($N_l=N$).}
    \label{feature2}
\end{figure*}
\subsection{Ablation Study}
\subsubsection{Analyzing the role of deep network components and semi-supervised learning} In this section, we examine the effectiveness of each component of the proposed SSDA architecture by reporting accuracy and F1 score values on both PhysioNet and BCI \uppercase\expandafter{\romannumeral4} 2a datasets. Table~\ref{ablation} summarizes the results.

CNN, LSTM, and Attention methods indicate a single layer of each individual model which are responsible for extracting the spatial, temporal, and attentive features, respectively. CNN-Attention and LSTM-Attention investigate the attentive information on top of CNN encoding and temporal dynamics, respectively.

In the PhysioNet dataset, the CNN-Attention and CNN-LSTM models achieve the highest accuracy and F1 score among the baseline models. However, the proposed SSDA method outperforms all the baseline models with a significant margin, achieving an accuracy of $0.83$ and an F1 score of $0.83$. In the BCI \uppercase\expandafter{\romannumeral4} 2a dataset, the CNN-LSTM model achieves the highest accuracy and F1 score among the baseline models. However, the proposed SSDA method outperforms all the baseline models, achieving an accuracy of $0.61$ and an F1 score of $0.59$. 

Comparing CNN, LSTM, and Attention models individually with combined models shows the importance of the combination of spatial and temporal encoding models. Our proposed SSDA that relies on spatial encoding, temporal dynamics, and attentive information represents not only the positive effect of each deep network backbones but also the effectiveness of the columnar structure of CST-AE. CST-AE consists of columns of CNN-RNN which helps with the performance improvement as shown in Table~\ref{ablation}.

To study the semi-supervised learning performance using various components of the deep network architecture, we apply the same limited labeled training samples to the aforementioned methods. As presented in Table~\ref{semisupervised}, with  $10\%$ labeled training samples, the proposed SSDA gives a performance higher than the chance level due to the semi-supervised learning process. These results indicate that the proposed SSDA outperforms other methods by attaining an accuracy of $0.78\pm0.03$ and an F1 score of $0.77\pm0.03$ on the PhysioNet dataset, as well as an accuracy of $0.48\pm0.05$ and an F1 score of $0.45\pm0.04$ on the BCI \uppercase\expandafter{\romannumeral4} 2a dataset. This demonstrates the effectiveness of the proposed SSDA method in leveraging both labeled and unlabeled data to enhance classification performance. 

On the other hand, the results also show that traditional deep learning models such as CNN, LSTM, and their combinations did not perform well on both datasets, especially on the BCI \uppercase\expandafter{\romannumeral4} 2a dataset, which is a challenging dataset due to the small number of labeled samples and more number of the classes. The Attention mechanism also did not improve the classification performance significantly. These results emphasize the importance of utilizing semi-supervised learning methods, such as the proposed SSDA, to improve the performance of deep learning models, especially on datasets with limited labeled data.

Also, by comparing the results in Table~\ref{semisupervised} and Table~\ref{ablation}, the positive effect of our full SSDA architecture on semi-supervised learning in four-class classification performance is vivid since leaving out some of the components of the architecture results in significant performance drops in the case of limited training samples.  
\subsubsection{The effect of center loss on the learned features}
As shown in Table~\ref{ablation}, the SSDA method outperforms other baselines. One major reason for this is the loss function, $\mathcal{L}$, used for training SSDA. To explore the effectiveness of optimizing the model with $\mathcal{L}$, we present the learned representations of the last layer of the network in Figs.~\ref{feature} and~\ref{feature2}, which show the distribution of the learned features in the last FC layer with and without considering the center loss in the classification part of the proposed network during the training phase for PhysioNet and BCI \uppercase\expandafter{\romannumeral4} 2a datasets, respectively. From Figs.~\ref{feature} and~\ref{feature2}, we can observe that the learned features under $\mathcal{L}$ have clearer boundaries between two and four MI classes compared to the ones that do not have a center loss in the defined loss function.
\subsection{Discussion}
\label{ssec:discussion}
In this study, we introduce a novel subject-independent approach for EEG-based MI tasks. The use of automated EEG-based motor imagery is crucial in neuroscience and brain-computer interfaces because it helps improve assistive technology and neurorehabilitation. It does this by reading people's thoughts from their brain signals, making it easier for them to control devices and systems effectively. Our model is composed of a deep unsupervised CST-AE in conjunction with a supervised deep classifier, which works well even when there's not much labeled data. It improves brain-computer interfaces and has potential in various applications.

For the $N_l=N$ scenario, as shown in Table~\ref{comparison}, the proposed SSDA reaches $0.83\pm0.03$ and $0.61\pm0.08$ classification accuracy for PhysioNet and BCI \uppercase\expandafter{\romannumeral4} 2a datasets, respectively, which outperforms the state-of-the-art works. The major reason that our proposed deep architecture performs better than the traditional FBCSP approach is its ability to learn high-level features from complex EEG data. This also removes the need to find suitable features for a specific domain. 
Comparing our work with other deep learning models reveals several key strengths. Firstly, the high degree of similarity between the original and reconstructed data indicates that our model successfully captures and retains relevant spatio-temporal patterns from the input EEG signals that are crucial for reconstruction \cite{nejedly2023utilization, li2020latent}. Secondly, CST-AE allows the utilization of different spatio-temporal windows to learn latent representations. This eliminates the challenge of determining the best kernel size, filter size, and the number of hidden states for CNN and LSTM architectures. Thirdly, unlike other deep learning algorithms, our DS approach facilitates learning representations in a lower dimension while maintaining discriminative ability. This fidelity in representation is especially crucial for subject-independent tasks, indicating the model's capacity to extract and generalize features indicative of task-related neural activities. Fourthly, $\mathcal{L}_c$  helps to optimize the problem by minimizing intra-class variation which improves the training process. Since the proposed approach optimizes both parts of the model in an end-to-end fashion, the classification goal influences the optimization of CST-AE, ensuring that the latent features extracted are relevant to the MI task. Despite the positive aspects associated with using these loss functions as demonstrated in the presented results, it is acknowledged that the computation of the loss may increase the training time, as a function of the amount of data. However, given that our method performs effectively with a small amount of labeled data, the time spent by the supervised components is reduced. In essence, this represents a minor trade-off between performance and time complexity.

One primary motivation behind the presented framework is addressing a prevalent challenge in EEG analysis: the scarcity of labeled training data ($N_l\ll N$). As presented in Table~\ref{sample} and considering $0.50$ and $0.25$ as chance levels for two class and four class classification scenarios, the results indicate that only $10\%$ labeled training samples reach a performance much higher than the chance level.  When analyzing the confusion matrices shown in Figs~\ref{confusion1} and~\ref{confusion2}, we find that even with only $10\%$ of labeled training samples, SSDA can still detect all four classes with significantly higher accuracy than the chance level. Moreover, training the proposed network with only $30\%$ of the labeled training samples outperforms most of the supervised state-of-the-art methods listed in Table~\ref{comparison}.

Statistical analysis is crucial to confirm the significance of the experimental findings when $N_l \ll N$ \cite{zhang2018learning}. To achieve this, we employ the Wilcoxon signed-rank test for each of the models considered in the ablation study alongside the proposed method. The results are presented in Table~\ref{pvalue}. Considering a p-value threshold of $0.05$ as the significance level, the results show that almost all of the performance differences between the proposed method and other methods considered in the ablation study are statistically significant.

In all experiments presented in this paper, we used raw EEG measurements without any preprocessing. Accordingly, the results presented demonstrate the performance of our approach on noisy data involving potential outliers collected within the context of particular motor imagery experimental paradigms. Future work could examine the robustness of our approach to other types of perturbations, possibly including adversarial examples. For further research directions, we can explore the performance of the proposed SSDA on different BCI modalities, such as P300, which could benefit from the CST-AE structure to find the attentive temporal dynamics. Additionally, SSDA's capability to train well with a small number of training samples could be tested on EEG-based emotion recognition, which is a complex BCI task that requires expert labeling.
\subsection{Limitations}
Even though the proposed study has introduced a novel and enhanced framework that surpasses the performance of previous methods, it still exhibits specific limitations. Firstly, there is the issue of the DS loss. The DS loss is defined among every pair of samples, leading to an increase in computation time. To mitigate this complexity, we can employ a random selection of pairs, thereby reducing the training time. Secondly, the use of fully connected layers could result in an increased number of parameters. To address this issue, future work could reduce the number of model parameters by incorporating global pooling layers while ensuring the maintenance of high motor imagery EEG decoding accuracy.
\section{Conclusion}
\label{sec: conclusion}
In this paper, we proposed a novel semi-supervised deep architecture to improve the subject-independent MI classification task. The proposed method consists of two parts: an unsupervised component and a supervised component. The unsupervised part, the CST-AE, extracts latent features by maximizing the similarity between original and reconstructed EEG data. The supervised part learns a classifier based on labeled samples using the latent features obtained from the unsupervised part. Additionally, center loss is employed to minimize the embedding space distance within each class. Our experimental results on two publicly available benchmarks, one with two-class and the other with four-class MI tasks, show superior performance compared to state-of-the-art works. Additionally, the distribution of the learned representations also demonstrates the positive effect of using center loss along with the classification loss. We also demonstrate that even a small portion of labeled training data samples can lead to effective classification performance due to the unsupervised part of SSDA. This work has the potential to reduce the need for a large number of labels in EEG tasks and eliminate the calibration stage for each subject.

\bibliographystyle{IEEEbibvv}
\bibliography{refs}

\end{document}